\documentclass[letterpaper, 10 pt, conference]{ieeeconf}  % Comment this line out if you need a4paper

\IEEEoverridecommandlockouts                              % This command is only needed if 
\usepackage{amsmath,graphicx}
\usepackage{verbatim}
\usepackage{subfig}
\usepackage{caption}
\usepackage{dblfloatfix}
\usepackage{balance}

\usepackage[noadjust]{cite}

\title{\LARGE \bf
Estimating Respiratory Rate From Breath Audio Obtained Through Wearable Microphones}
 
\author{Agni Kumar, Vikramjit Mitra, Carolyn Oliver, Adeeti Ullal, Matt Biddulph, Irida Mance% <-this % stops a space

\thanks{Apple, Cupertino, California, USA. Emails: \tt\small\{agni@apple.com, vmitra@apple.com, carolyno@apple.com\}}}

\begin{document}

\maketitle
\thispagestyle{empty}
\pagestyle{empty}

%%%%%%%%%%%%%%%%%%%%%%%%%%%%%%%%%%%%%%
\begin{abstract}
\textit{Respiratory rate} (RR) is a clinical metric used to assess overall health and physical fitness. An individual's RR can change from their baseline due to chronic illness symptoms (e.g., asthma, congestive heart failure), acute illness (e.g., breathlessness due to infection), and over the course of the day due to physical exhaustion during heightened exertion. Remote estimation of RR can offer a cost-effective method to track disease progression and cardio-respiratory fitness over time. This work investigates a model-driven approach to estimate RR from short audio segments obtained after physical exertion in healthy adults. Data was collected from 21 individuals using microphone-enabled, near-field headphones before, during, and after strenuous exercise. RR was manually annotated by counting perceived inhalations and exhalations. A multi-task Long-Short Term Memory (LSTM) network with convolutional layers was implemented to process mel-filterbank energies, estimate RR in varying background noise conditions, and predict heavy breathing, indicated by an RR of more than 25 breaths per minute. The multi-task model performs both classification and regression tasks and leverages a mixture of loss functions. It was observed that RR can be estimated with a concordance correlation coefficient (\textit{CCC}) of 0.76 and a mean squared error (MSE) of 0.2, demonstrating that audio can be a viable signal for approximating RR.
\newline

\indent \textit{Clinical relevance}—The subject technology facilitates the use of accessible, aesthetically acceptable wearable headphones to provide a technologically efficient and cost-effective method to estimate respiratory rate and track cardio-respiratory fitness over time.
\end{abstract}

%%%%%%%%%%%%%%%%%%%%%%%%%%%%%%%%%%%%%%
\section{Introduction}
\label{sec:intro}
Breathlessness, or dyspnea, is a common symptom in many acute and chronic clinical conditions. Acute breathlessness often occurs during an asthmatic episode \cite{barbaro} or heart attack \cite{nakanishi}, while chronic breathlessness is frequently a symptom of low cardiovascular fitness and obesity \cite{bernhardt,jensen}, chronic obstructive pulmonary disease (COPD) \cite{celli}, and congestive heart failure (CHF) \cite{hendren, pang}. Breathlessness on exertion is also a strong independent predictor of mortality \cite{pan} and is a commonly-used clinical metric for assessing and monitoring disease progression. The primary classifications of heart failure (NYHA Class I-IV) are defined in terms of breathlessness, either occurring at rest or during normal levels of physical activity \cite{criteria1994nomenclature}. 

Breathlessness scores as quantified by the Borg Dyspnea Scale, used to assess clinical severity of diseases such as peripheral artery disease (PAD) and other respiratory disorders, are subjective patient-reported measures \cite{borg}. Individuals with such conditions are typically required to interface with a healthcare provider in order for their symptoms to be recognized. Creating an objective method for breathlessness detection would lower the burden in identifying this symptom and may even alert healthcare providers to patients' underlying medical conditions long before disease progression would have been observed in a clinical setting. 

In this paper, we take the first step towards developing a breathlessness measurement tool by estimating respiratory rate (RR) on exertion in a healthy population using audio from wearable headphones. Given this focus, such a capability also offers a cost-effective method to track cardio-respiratory fitness over time. While sensors such as thermistors, respiratory gauge transducers, and acoustic sensors provide the most accurate estimation of a person’s breathing patterns, they are intrusive and may not be comfortable for everyday use. In contrast, wearable headphones are relatively economical, accessible, comfortable, and aesthetically acceptable. 

Previous work on breath sounds detection from audio has focused on the detection and categorization of particular breath sounds using breath characteristics to distinguish between healthy and abnormal breath sounds \cite{li2017design, castro2014real}, as well as RR estimation from both contact-based sensors to obtain tracheal sounds \cite{sierra2006comparison, sierra2004monitoring} and non-contact-based sensors, like smartphones, to acquire nasal breath recordings \cite{ren2015fine}. The goal of this study is to validate whether versatile non-contact sensors, such as wearable near-field microphones, can provide incoming audio data sufficient to distinguish between normal and heavy breathing and also estimate RR by the sensing of breath sound patterns. Though we focus on discerning RR in the context of fitness activity and not clinical breathlessness detection specifically, findings could in turn be used for medical applications.

%%%%%%%%%%%%%%%%%%%%%%%%%%%%%%%%%%%%%%
\section{Data}
\label{sec:Data}
We chose to focus on the exercise context to obtain breath samples of varying intensities. Data was collected from 21 healthy individuals from both indoor and outdoor environments. Participants spanned the ages of 22 to 60 and were split fairly evenly between genders. Several participants provided six pulse rate measurements per audio sample submitted, each of which spanned a six-minute active period. 

All data was recorded using microphone-enabled, near-range headphones, specifically Apple's AirPods. These particular wearables were selected because they are owned by millions and utilized in a wide array of contexts, from speaking on the phone to listening to music during exercise. Unlike prior studies in which the acoustic environment and background noise were artificially altered \cite{ren2015fine}, our study did not place artificial acoustic constraints on the data collection setup. 

Each trial of the data collection asked participants to record four one-minute audio clips before, during, immediately after, and while cooling down following completion of a nine-minute workout session, in which six minutes involved physical exercise. Workout types were selected to induce heavy breathing, with the goal of doubling participants' resting heart rates at the peak of physical exertion.

\begin{itemize}
    \item [1.] Before workout (minute 1)
    \item [2.] During the workout, towards the end (minute 5)
    \item [3.] Immediately after the workout (minute 6)
    \item [4.] While cooling down (minute 8)
\end{itemize}

\begin{figure}[h]
  \centering
  \includegraphics[width=0.47\textwidth]{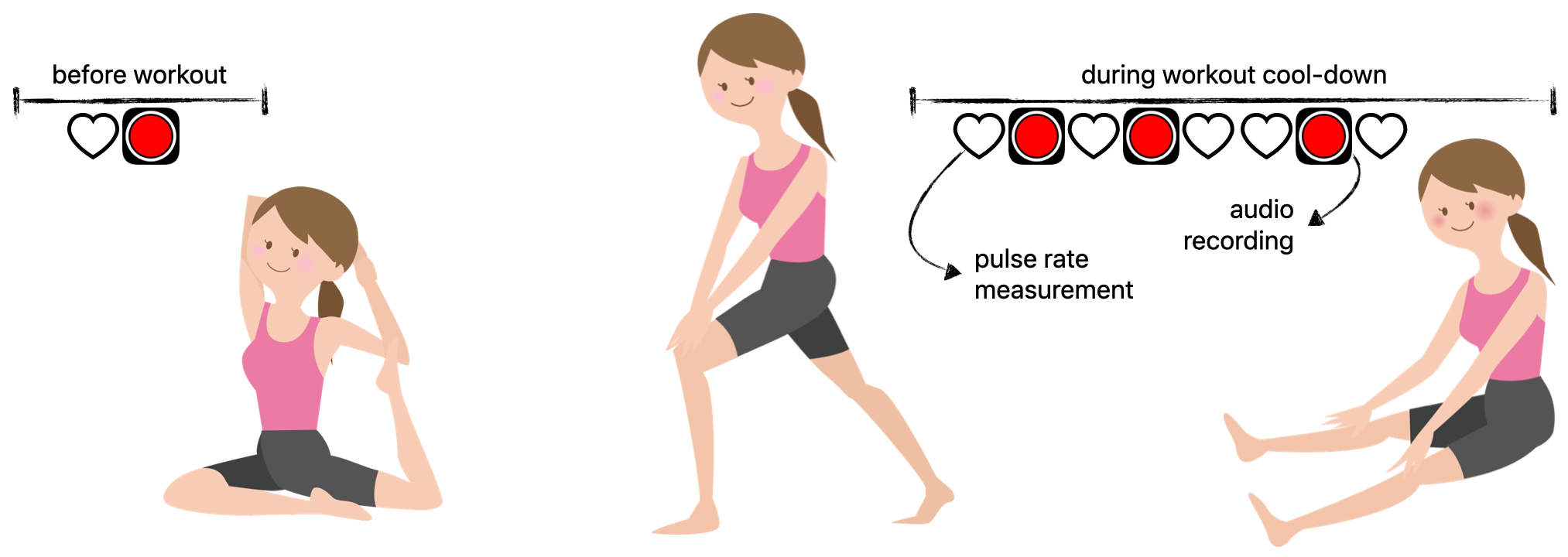}
  \caption{One trial of data collection included four recording types corresponding to distinct workout stages.}
  \label{fig:data_protocol}
  \raggedleft
\end{figure}

Participants were also asked to note pulse rates in beats per minute (BPM) as obtained from an Apple Watch at six points in the exercise session: before recording an audio clip of type \boxed{1}, before recording an audio clip of type \boxed{2} (minute 5), and before and after recording an audio clip of each of type \boxed{3} (minutes 6 and 7) and type \boxed{4} (minutes 8 and 9). The data collection protocol is summarized in Figure \ref{fig:data_protocol}. % \ref{fig:data_protocol_2} for trial totals

\setlength{\textfloatsep}{2pt plus 1.0pt minus 1.0pt}

All audio sessions were segmented into randomly-selected lengths between 4-7 seconds, to increase the probability that a segment would contain at least one breath cycle. 3003 segments were manually annotated for RR. The annotation process involved counting the number of inhale-exhale cycles that could be heard in each audio sample and dividing the breath count by the clip duration in minutes to achieve respiratory rate measures in breaths per minute. 

299 segments of type \boxed{1}, 457 of type \boxed{2}, 762 of type \boxed{3}, and 285 of type \boxed{4}, along with 1200 randomly chosen segments, were annotated and served as ground truth values. The perceptual annotation was motivated by the fact that perceived audio respiration events should be represented well by the acoustic features, which in turn would facilitate the model training. In cases of high background noise, audio samples could be rendered imperceptible to the human ear and result in distorted acoustic features. 

In the spectrogram in Figure 2(a), indicative of normal breathing, two prominent inhalation signals are observed with no harmonic structure. The intense exercise sample in Figure 2(b), associated with heavy breathing, shows more frequent energy bursts but also lacks harmonic structure, indicating both a higher RR and greater background noise. These observations support the use of temporal spectral representations in distinguishing between normal and heavy breathing, which prompted using temporal convolution and recurrent layers in the model described in Section \ref{sec:system_desc}.

\begin{figure}[h]
\label{fig:spectrograms}
\subfloat[normal breathing]{\includegraphics[width=0.47\textwidth]{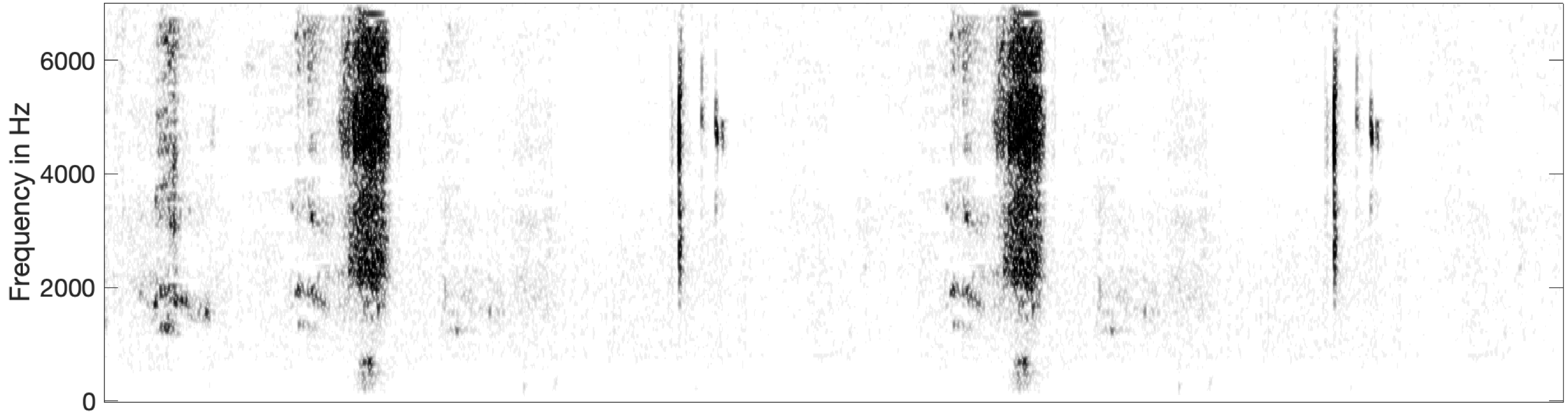}}
\newline
\subfloat[heavy breathing]{\includegraphics[width=0.47\textwidth]{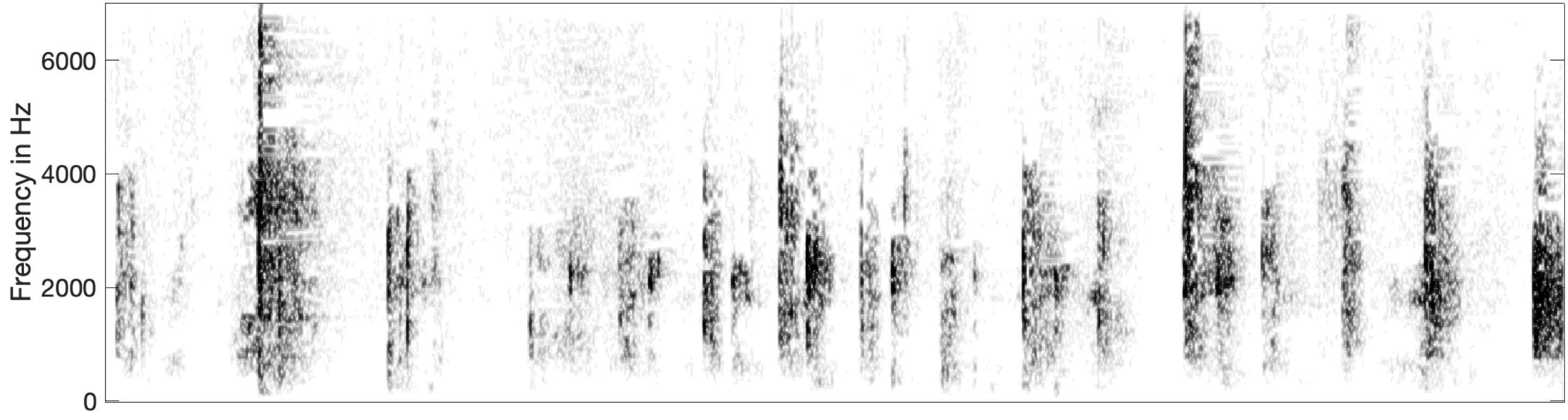}}
\caption{Breath audio spectrograms}
\end{figure}

In total, over 3000 annotated audio signals were analyzed. We are not aware of any audio data resources that have perceptually graded RR labels that are bigger in volume than that collected for this work. Analyses by breathing intensity and gender are thus provided in Section \ref{sec:results}.

%%%%%%%%%%%%%%%%%%%%%%%%%%%%%%%%%%%%%%
\section{Analysis} 
\label{sec:Analysis} 
\begin{figure}[h]
  \centering
  \includegraphics[width=0.47\textwidth]{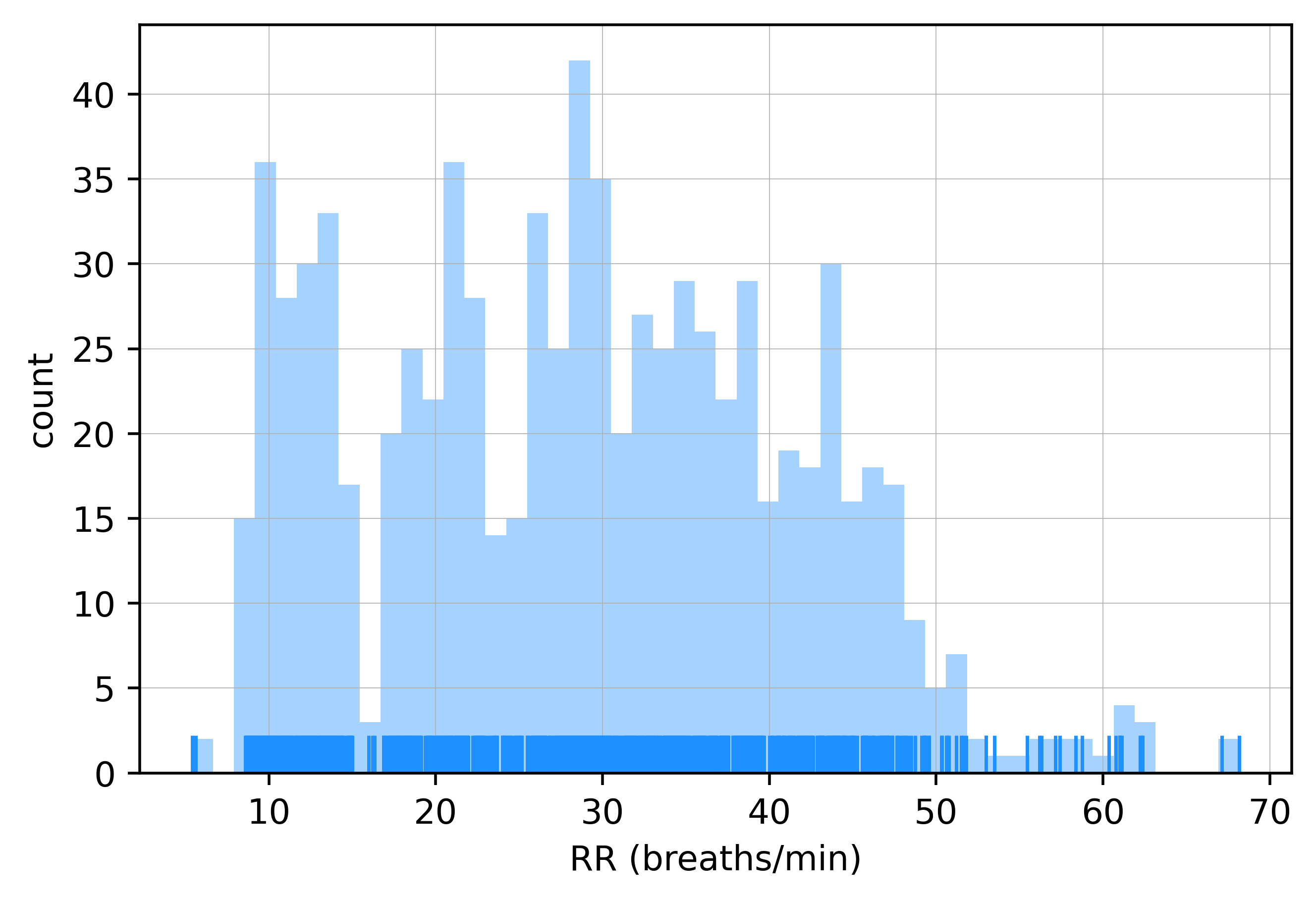}
  \caption{Distribution of annotated respiratory rates}
  \label{fig:rate_dist}
  \raggedleft
\end{figure}

We observe several interesting relations in the data. In Figure \ref{fig:rate_dist}, we note a right-tailed distribution of RR, likely due to some participants having not exerted themselves to the level requested, perhaps because it was not achievable during a six-minute period with particular workout types. In a future data collection, efforts will be made to either upsample data yielding higher respiratory rates or extend the workout duration to increase the chances of observing heavy breathing. 

\begin{figure*}[ht!]
  \centering
  \captionsetup{justification=centering}
  \includegraphics[width=0.98\textwidth]{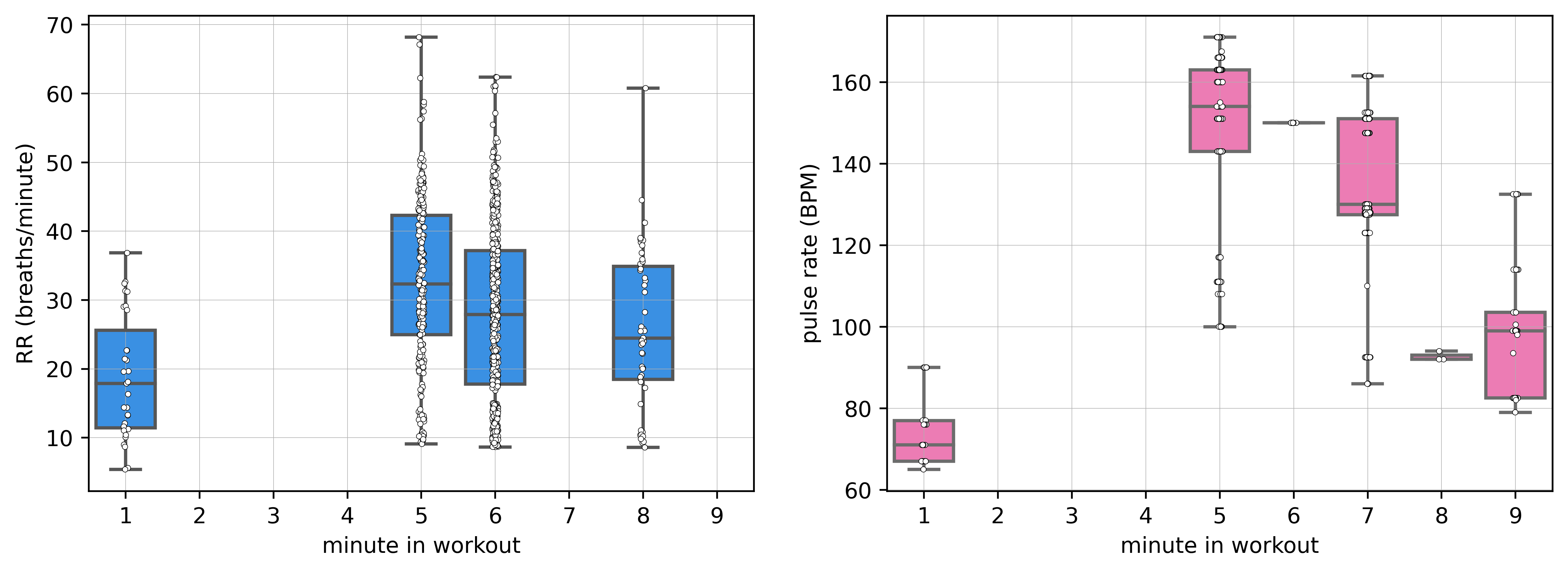}
  \caption{Respiratory and pulse rates over workout stages}
  \label{fig:breath_pulse}
\end{figure*}

Participants' exercises varied in intensity and included running, biking, HIIT (high-intensity interval training), calisthenics, fast walking, and walking up an incline. 60\% of audio samples were recorded indoors while 40\% were taken outdoors. It was observed that most of the data that was not entirely noisy, i.e., breath-only or a mix of breath and noise, comprised samples from during and immediately after workouts, when breathing was heaviest. Moreover, participants' noise contributions differed significantly, likely due to differing workout environments and home exercise equipment. Data from users who completed repeated exercises in close time proximity tended to exhibit higher-than-normal resting RR and pulse rates, since breaks taken in between were sometimes not long enough to reestablish a completely non-exerted state.

In Figure \ref{fig:breath_pulse}, we visualize how participants' respiratory and pulse rates vary over time across workout stages, represented by individual blue boxes. As expected, the median pulse rate is lower for stages \boxed{1} and \boxed{4} than for \boxed{2} and \boxed{3}, and that the pulse rate associated with \boxed{2} is higher than that associated with \boxed{3}. RR varies in a similar manner, and the high RR values achieved align with those collected in similar studies \cite{nicolo2017respiratory}. Each overlaid white circle is associated with a single audio segment. The numbers of white circles differed between minutes because participants could choose to opt-in or opt-out of submitting audio samples at any point in time, as the study was designed to ensure high privacy. 

%%%%%%%%%%%%%%%%%%%%%%%%%%%%%%%%%%%%%%
\section{System Description}
\label{sec:system_desc}
We investigated mel-filterbank energy (MFB) acoustic features to parameterize the audio streams \cite{heo2019end} and trained an LSTM network and a time-convolutional LSTM (TC-LSTM) network using multi-task learning (MTL) \cite{kendall2018multi}. Our study presents, for the first time, the use of filterbank energies processed by time-convolutional layers. We hypothesize that such a setup is effective as it captures both signal envelope-level information as well as longer time contexts to better assess respiration information. The learning network, an end-to-end model, is not standard in that it simultaneously encompasses both regression and classification tasks. Its architecture was motivated by the nature of the specific data collected, described in Section \ref{sec:Data}.

\subsection{Acoustic Features}
We examined standard 40-dimensional MFB features with upper and lower cut-off frequencies of 0 and 7500 Hz. We investigated MFBs with both more than and less than 40 filteranks, and observed that 40 filterbanks offered the best performance on the held-out validation set. 

This frequency range was chosen as exhalation signals contain wideband information, with \cite{niu2019novel} noting that frequencies between 20 to 6000 Hz can encompass useful information. From our analysis, we observed that nasal and oral exhalation had very different spectral characteristics, with the former having low frequency band-pass characteristics and the latter having more energy in high-pass regions. (Similarly, when comparing nasal and oral snores, \cite{mikami2012classification} found nasal snores to exhibit frequencies around 500 Hz, while oral snores were associated with wide-band spectra ranging from 500 Hz to 1000 Hz.) Given that this study considers data collected from subjects after intensive workouts that may result in more oral breathing due to exhaustion, we hypothesized that energies in high frequency spectrums could be quite useful. Moreover, as one of the multi-task learned model’s objective functions relates to noise detection, the selected MFB feature frequency range is fitting.
%We also investigated 40-dimensional modulation spectra energy features (MOD40) \cite{mitra2012normalized} that enabled exploration of feature fusion experiments presented in Section \ref{sec:results}.

\begin{figure}[h]
  \centering
  \includegraphics[width=0.47\textwidth]{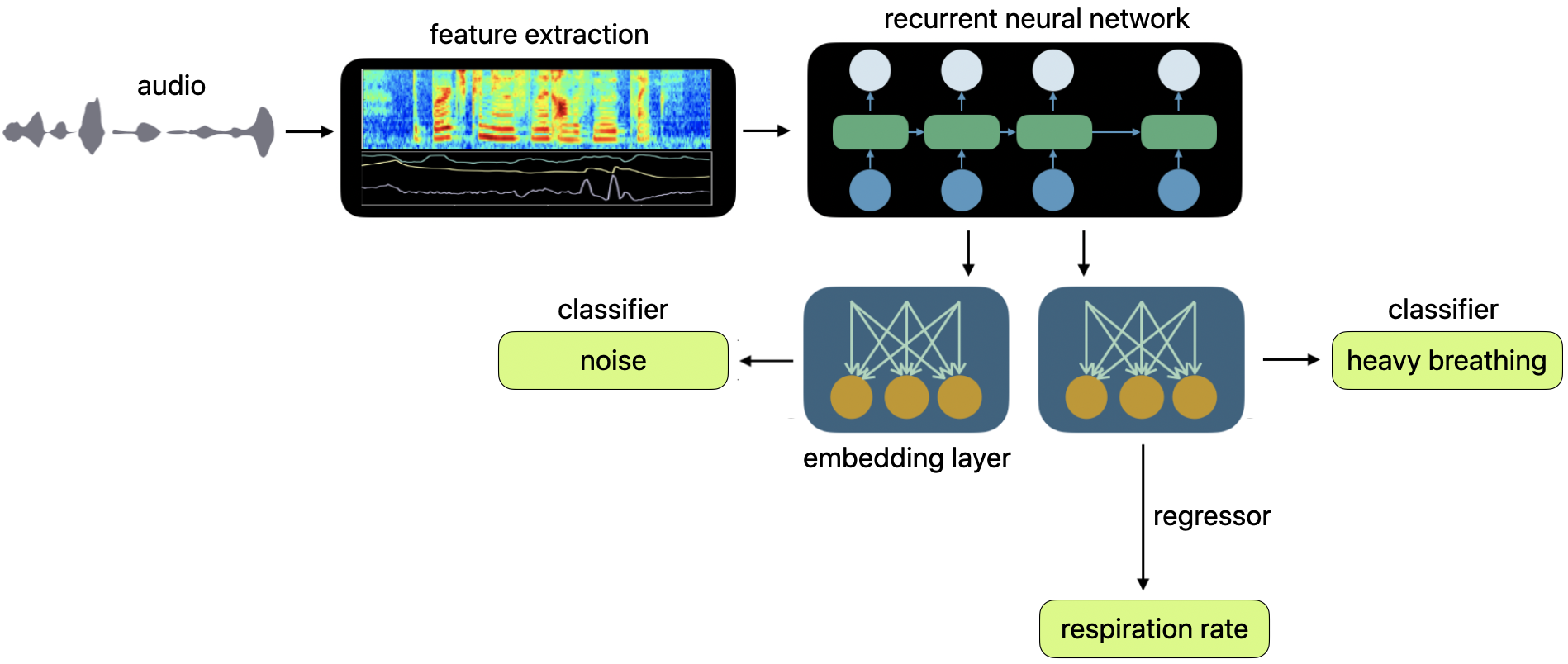}
  \caption{Multi-task learning architecture with LSTM}
  \label{fig:model_diagram}
  \raggedleft
\end{figure}

\subsection{Acoustic Modeling}
We investigated single layer LSTMs with 16, 32, and 64 neurons in the recurrent and embedding layers. The network had three outputs: 
\begin{itemize}
    \item A 2-dimensional output corresponding to RR and respiration count (RC) estimates
    \item A 3-dimensional output reflecting the classes of no breathing, normal breathing (RR between 5 to 30 breaths per minute), and heavy breathing (RR greater than 30 breaths per minute)
    \item A 2-dimensional output corresponding to noise or noiseless cases
\end{itemize}

The model, depicted in Figure \ref{fig:model_diagram}, was trained with multiple objective functions as a multi-task learning (MTL) network, where the tasks were RR estimation, heavy breathing detection, and noise detection, represented by the three outputs mentioned above. As the audio segments used to train the model had varying lengths, meaning that two segments of differing durations could have the same RR but different RCs, RC  was included as a target in the multi-task objective function to ensure the model's ability to generalize across temporal durations and also learn correlations between RC and RR. 

The individual losses from each task are given below, where concordance correlation coefficient (\textit{CCC}) loss is used on the RR and RC outputs and weighted cross-entropy (CE) loss is used on the breath and noise classification tasks. The interpretation of \textit{CCC} used here is similar to that of Pearson’s product moment correlation (\textit{PPMC}), where \textit{CCC} ranges between $-1$ to $+1$ and a higher value indicates a better fit \cite{bland1986statistical}. \textit{CCC} was chosen as a performance metric based on findings from the literature showing that it captures both aspects of the correlation and the error between target and estimated scores \cite{lawrence1989concordance, kowtha2020detecting}.
\begin{equation*}
\begin{aligned}
{CCC_{\text{cost}}:= \alpha CCC_{\text{RR}}+(1-\alpha)CCC_{\text{RC}}} \\
{CE(x)_{\text{breath}}:= w_{\text{breath}}(-x_{\text{breath}} + \log \sum_{i}\exp(x[i])}) \\
{CE(y)_{\text{noise}}:= w_{\text{noise}}(-x_{\text{noise}} + \log \sum_{j}\exp(y[j] )}) 
\end{aligned}
\end{equation*}

Additionally, a focal loss term \cite{linfocal,gil2020balance} was used for the breath detection task, and a convex mixture of all the losses after dynamic weight averaging \cite{liu2019end}, with weighting factor $\lambda$, was used as the MTL loss to train the network shown in Figure \ref{fig:model_diagram}.
\begin{equation*}
\begin{aligned}
MTL_{\text{loss}} &= \beta \cdot \lambda_{CCC} \cdot CCC_{\text{\text{cost}}} \\
&+ \gamma \cdot \lambda_{CE_{\text{breath}}} \cdot CE(x)_{\text{breath}} \\
&+ \kappa \cdot \lambda_{CE_{\text{noise}}} \cdot CE(y)_{\text{noise}} \\
&+ (1-\beta-\gamma-\kappa) \cdot \lambda_{FL_{\text{breath}}} \cdot FL(x)_{\text{breath}} \\ 
\end{aligned}
\end{equation*}

The \textit{CCC} for each of the RR and RC outputs is below, where ${\mu _{x}}$ and ${\mu _{y}}$ are the means, ${\sigma _{x}^{2}}$ and ${\sigma _{y}^{2}}$ are the corresponding variances for the estimated and ground truth variables, and ${\rho}$ describes the correlation between these two variables.
\begin{equation*}
\begin{aligned}
CCC &= \frac {2\rho \sigma_x \sigma_y}{\sigma_x^2+\sigma_y^2 +(\mu_x-\mu_y)^2 }
\end{aligned}
\end{equation*}

The model was trained with a mini-batch size of 64 using an Adam optimizer, with a learning rate of 0.01 and a momentum of 0.9. For all the model training steps, early stopping was allowed based on cross-validation error. The embedding layer between the breath classification and RR and RC estimation tasks was shared. The noise classification task had a separate embedding layer, which was a fourth of the size of the breath embedding layer.

We also investigated a TC-LSTM network, where a convolution operation with a filter size of 3 was performed across time. The number of convolution filters was same as the number of input feature dimensions. The feature maps from the convolutional layers were fed as inputs to the succeeding LSTM network. 

%, as shown in Figure \ref{fig:tc_model_diagram}.

% \begin{figure}[h]
%   \centering
%   \includegraphics[width=0.47\textwidth]{figures/model_diagram_baseline.png}
%   \caption{Multi-task learning architecture with LSTM}
%   \label{fig:model_diagram}
%   \raggedleft
% \end{figure}

%\begin{figure}[h]
%  \centering
%  \includegraphics[width=0.47\textwidth]{figures/model_diagram_tc.png}
%  \caption{Multi-task learning architecture with TC-LSTM}
%  \label{fig:tc_model_diagram}
%  \raggedleft
%\end{figure}
During training, the \textit{CCC} for RR estimation on a held-out validation set was used to select the best epoch, and the model from that epoch was used to obtain the performance on the held-out evaluation set. Results on the evaluation set are provided in terms of both \textit{CCC} and mean squared error (MSE).

%%%%%%%%%%%%%%%%%%%%%%%%%%%%%%%%%%%%%%
\section{Results}
\label{sec:results}

We investigated model size by exploring LSTM models with 16, 32, and 64 neurons in the recurrent and embedding layers. The evaluation metrics used were $F_1$ scores for the classification task and \textit{CCC} for the regression task. 
% \begin{figure*}[!b]
% \centering
% \captionsetup{justification=centering}
% \includegraphics[width=.98\textwidth]{figures/mse_comparisons_labels.png}
% \caption{Performance (MSE) comparisons at different RR ranges for female and male participants \\ in the test set, across the LSTM and TC-LSTM models before and after data augmentation}
% \label{comparisons}
% \end{figure*}

Table \ref{tab:table1} shows the variation in performance by model size, where the model with 32 neurons in the LSTM layer, 32 neurons in the breath embedding layer, and 8 neurons in the noise embedding layer was found to perform the best on both the validation and evaluation sets. The model with 64 neurons shows some degree of overfitting, where the performance gap between the validation and evaluation set was larger than that of the model with 32 neurons. This could be a consequence of the data volume limitation, as larger datasets may enable using models with more parameters.

\begin{table}[h]
%\small
\centering
\caption{\textit{CCC}, \textit{PPMCC} ($\rho$) for RR estimations, and $F_1$ scores for breath and noise classification tasks for the validation and evaluation sets, from the MTL-LSTM network}
  \begin{tabular}{lcccc}
    % \hline
    Neurons & {$CCC_{\text{RR}}$} & {$\rho_{\text{RR}}$} & {$F_{1,\text{breath}}$} & {$F_{1,\text{noise}}$} \\
    \hline     
    \textit{Validation} &  &  &  &  \\
    $16$ & 0.79 & 0.85 & 57.55 & 88.82 \\
    $32$ & \textbf{0.88} & \textbf{0.88} & \textbf{64.76} & 89.61 \\
    $64$ & 0.85 & 0.85 & 56.53 & \textbf{93.65} \\
    \hline
    \textit{Evaluation} &  &  &  &  \\
    $16$ & 0.59 & 0.65 & 49.20 & 66.91 \\
    $32$ & \textbf{0.73} & \textbf{0.73} & \textbf{66.33} & \textbf{76.81} \\
    $64$ & 0.62 & 0.62 & 52.82 & 65.04 \\
    \hline
  \end{tabular}
\label{tab:table1}
\end{table}

Next, we explored a time-convolution layer in the acoustic model and modulation features as an alternative feature representation, and whether such selections could facilitate better detection of respiration from audio. Table \ref{tab:table2} presents the results from these experiments, for which there were 32 neurons in the LSTM and breath embedding layers. RR was estimated with a \textit{CCC} as high as 0.76 and a detection accuracy of breathing at 66\%. 

\begin{table}[h]
\centering
\caption{\textit{CCC}, \textit{PPMCC} ($\rho$), and MSE scores for RR estimation, and $F_1$ scores for breath classification tasks for the evaluation set}
  \begin{tabular}{lcccc}
    % \hline
    Model & {$CCC_{\text{RR}}$} & {$\rho_{\text{RR}}$} & {$MSE_{\text{RR}}$} & {$F_{1,\text{breath}}$} \\
    \hline     
    LSTM & 0.73 & 0.73 & 0.32 & 66.33 \\
    %MOD40 & LSTM & 0.69 & 0.67 & 71.95 \\
    TC-LSTM & 0.76 & 0.78 & 0.31 & 63.00  \\
    %MOD40 & TC-LSTM & 0.70 & 0.68 & 56.00   \\
    %MFB40 \\
    %+MOD40 & TC-LSTM & & &  \\
    \hline
  \end{tabular}
\label{tab:table2}
\end{table}

\begin{figure*}[t] % top placement
\centering
\captionsetup{justification=centering}
\includegraphics[width=.98\textwidth]{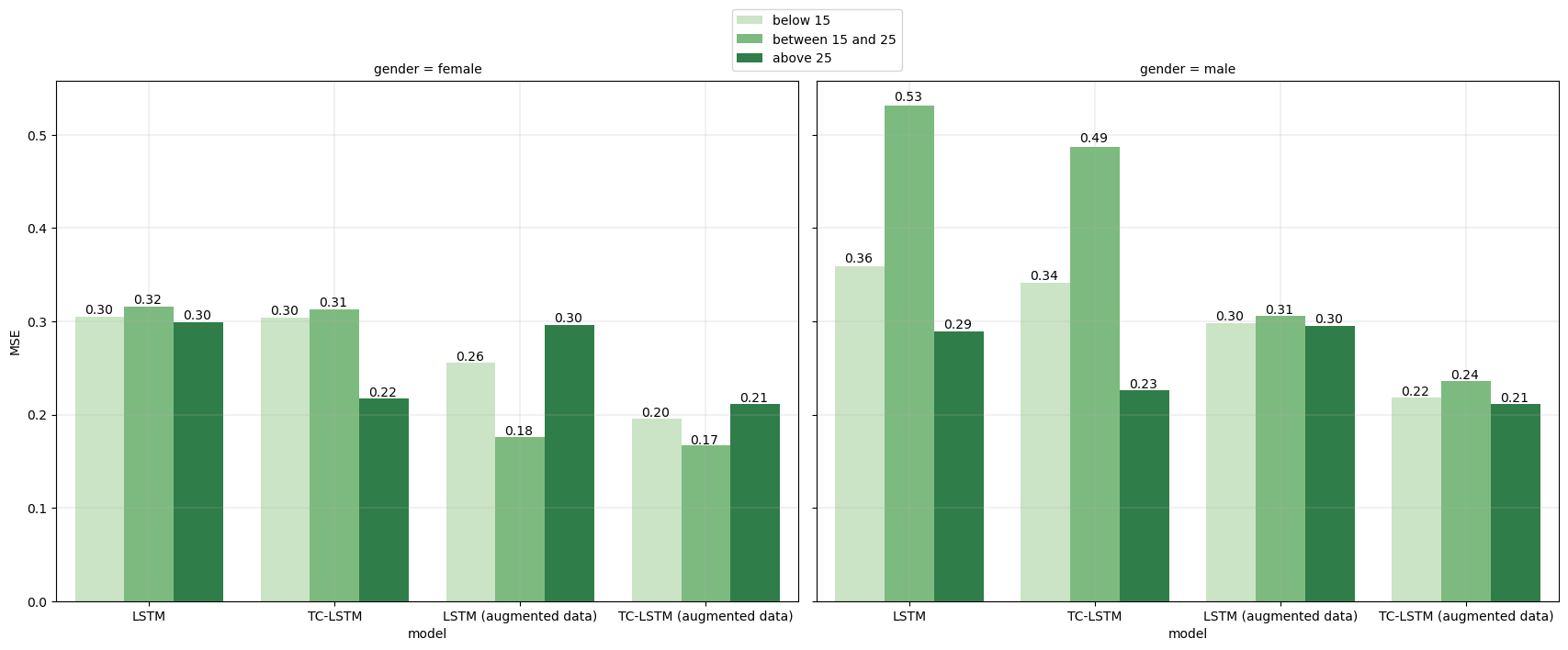}
\caption{Performance (MSE) comparisons for various RR ranges for female and male participants in the test set}
\label{comparisons}
\end{figure*}

% \begin{figure*}[!b] % bottom placement
% \centering
% \captionsetup{justification=centering}
% \includegraphics[width=.98\textwidth]{figures/mse_comparisons_labels.png}
% \caption{Performance (MSE) comparisons for various RR ranges for female and male participants in the test set}
% \label{comparisons}
% \end{figure*}

We then investigated how RR estimation varied for different RR ranges and how data augmentation could help improve the performance of the RR estimation model. Recall that the dataset used in this study included audio in both indoor and outdoor conditions. The outdoor data already contained natural ambient noise such as wind and traffic sounds, so the indoor data was further augmented by pseudo-stationary noise reflective of appliance sounds at various signal-to-noise ratios (SNRs) between 20 to 40 dBs. For each indoor data file, noise was added at three different SNR levels, each of which was selected from a uniform distribution between 10 to 20 dB, 20 to 30dB and 30 to 40dB, respectively. Data augmentation was applied only on the training partition. 

\begin{table}[h]
\centering
\caption{MSE values across the LSTM and TC-LSTM models before and after data augmentation across RR ranges}
  \begin{tabular}{lcccc}
    % \hline
    Model & below 15 & 15 to 25 & above 25 \\
    \hline     
    LSTM & 0.33 & 0.43 & 0.29 \\
    TC-LSTM & 0.32 & 0.42 & 0.22  \\
    LSTM (augumented data) & 0.28 & 0.24 & 0.29 \\
    TC-LSTM (augmented data) & \textbf{0.21} & \textbf{0.20} & \textbf{0.21} \\
    \hline
  \end{tabular}
\label{tab:table3}
\end{table}

Table \ref{tab:table3} shows how the performance of RR estimation in terms of MSE varied at low (less than 15 breaths per minute), medium (between 15 and 25 breaths per minute), and high (greater than 25 breaths per minute) RR rates. It is evident that data augmentation helped reduce the MSE for almost all the RR ranges. 
Moreover, as displayed in Figure \ref{comparisons} above, the TC-LSTM model performed better than the LSTM model both with and without data augmentation across female and male participants in the test set. Interestingly, estimates for female participants were almost always better than those for the male participants. It is possible that differing anatomical breathing apertures, age distributions, or ratios of indoor to outdoor samples collected between the two groups could have contributed to this observation.

%%%%%%%%%%%%%%%%%%%%%%%%%%%%%%%%%%%%%%
\section{Conclusions}
In this study, we investigated the feasibility of estimating respiratory rate from audio captured using wearable, near-field microphones. We found a time convolution LSTM network to be effective at generating respiratory rate estimates and robust against data sparsity. 

The work is unique in three main ways, in that it estimates respiratory rate from a wearable microphone under natural ambient conditions both indoors and outdoors, uses a model-driven approach to estimate respiratory rate directly from filterbank energies, and introduces situational awareness through multi-task learning so that the model could discern high SNR conditions from low ones. To the best of our knowledge, no prior study has investigated data collected from natural conditions from both indoor and outdoor background conditions, used perceptually graded data, and attempted to build an end-to-end system that can consume filterbank energies to directly predict respiratory rates and make heavy breathing classifications. Hence, we hope our proposed system will be used as a baseline for future studies on end-to-end respiratory rate estimation models. 

Results presented validate that RR can be estimated from audio captured using wearable microphones, enabling the detection of heavy breathing conditions and the monitoring of RR changes, a measure of cardio-respiratory fitness, over time. Data augmentation with simple acoustic distortion was demonstrated to be an effective tool to reduce error rates. The findings show promise for further development of a respiratory health tool with a larger study cohort.  
% \section*{Appendix}
% [INSERT]

%%%%%%%%%%%%%%%%%%%%%%%%%%%%%%%%%%%%%%
\section*{Acknowledgements}
We would like to thank Siddharth Khullar, Jamie Cheng, Suki Lee, Christopher Webb, Kate Neihaus, and many other colleagues for their comments, suggestions, and help at various stages of this effort. 

%%%%%%%%%%%%%%%%%%%%%%%%%%%%%%%%%%%%%%
\bibliographystyle{IEEEtran}
\balance
\bibliography{refs}
\end{document}